\documentstyle[amssymb]{article}

\begin{document}

\title{On the Equivalence Principle and gravitational and inertial mass
relation of classical charged particles}

\author{Mario Goto\\
Departamento de F\'{\i}sica,
Universidade Estadual de Londrina\\
86051-990, Londrina, PR, Brazil\\
(mgoto@uel.br)\\ \\
P. L. Natti\\
Departamento de Matem\'atica,
Universidade Estadual de Londrina\\
86051-990 Londrina, PR, Brazil\\
(plnatti@uel.br)\\ \\
E. R. Takano Natti\\
Pontif\'{\i}cia Universidade Cat\'olica do Paran\'a\\
Rua J\'oquei Clube, 458, 86067-000, Londrina, PR, Brazil\\
(erica.natti@pucpr.br)}

\maketitle

\begin{abstract}

We show that the locally constant force necessary to get a stable hyperbolic
motion regime for classical charged point particles, actually, is a 
combination of an applied external force and of the electromagnetic 
radiation reaction force.
It implies, as the strong Equivalence Principle is valid, that the passive 
gravitational mass of a charged point particle should be slight greater than 
its inertial mass. An interesting new feature that emerges from the 
unexpected behavior of the gravitational and inertial mass relation, for 
classical charged particles,  at very strong gravitational field, 
is the existence of a critical, particle dependent, 
gravitational field value that signs the validity domain of the strong 
Equivalence Principle. For electron and proton, these critical field 
values are $g_{c}\simeq 4.8\times 10^{31}m/s^{2}$ and $g_{c}\simeq 8.8\times
10^{34}m/s^{2}$, respectively.

\vskip 0.5cm


\end{abstract}

\newpage 

\section{Introduction}

The problem of the electromagnetic radiation reaction force on the charged 
particle dynamics, as given by the Lorentz-Abraham-Dirac (LAD) equation
\cite{Dirac}-\cite{Jackson},
has been a subject of active investigation. There are a lot of works about
this subject accumulated since the first
attempt was made by Dirac \cite{Dirac}.

There is now a renewed interest on this subject,
with works pointing to something new,
which should affect the validity
of the Weak Equivalence Principle at some circumstances 
\cite{Ginzburg}-\cite{Donoghue}.
Perhaps because the main experimental justification that led Einstein to
formulate the Equivalence Principle (EP), which is one of the foundations 
of his General Theory of Relativity, is the numerical
equality between inertial and gravitational mass, nowadays they are taken
quite as synonymous, so we have to be aware to avoid misleading conclusions.
According to Weinberg \cite{Steven}, we distinguish the Weak
Equivalence Principle (WEP) of the Strong Equivalence Principle (SEP). 
The Strong
Equivalence Principle postulates that at every space-time point in a
arbitrary gravitational field it is possible to choose a locally inertial
coordinate system such that, within a sufficiently small region of the
point in question, the laws of the nature take the same form as in
unaccelerated Cartesian coordinate systems in the absence of gravitation.
On the other hand, the Weak Equivalence Principle is nothing but a
restatement of the observed equality of gravitational and inertial mass.

About the verification of the WEP, 
there is a surprising richness in the variety of experimental 
techniques and choice of the test bodies which have been used 
so far. The equality of gravitational and inertial mass is in fact what 
the experiments, since the famous E\"otvos balance until recents 
experiments, actually measure. We show a brief review. 
The most obvious way to proof the WEP is to compare the motion
of two bodies during free fall. These experiments, limited by rather
short free falling periods, reached an accuracy of about 
$1$ part in $10^{-10}$ \cite{Galile}. The
Bremen drop tower experiments, using SQUID displacement sensors, will 
provide a much longer time for free fall, allowing to reach an accuracy 
of about $10^{-12}-10^{-13}$ \cite{Vodel,drop}. Torsion balance experiments have 
reached an accuracy of few parts in $10^{-13}$ \cite{torcao}. 
A planned experiment using 
a cryogenic balance claims an accuracy of $10^{-14}$ \cite{criog}. 
The most sensitive long-range measurements have used the Sun
as the source and Earth and Moon type test bodies. The lunar laser ranging 
(LLR) techniques reach an accuracy of $5\times 10^{-13}$ \cite{LLR1,LLR2}.  
On the other hand, future space experiments promise much better precision 
in this measurement. The MICROSCOPE mission \cite{Pradels} aims to test, 
on a microsatellite of the MYRIADE series developed by CNES/FRANCE, the WEP 
with a $10^{-15}$ accuracy. The Galileo Galilei-GG is a proposed experiment
in low orbit around the Earth aiming to test the WEP to the level of $1$ part
in $10^{-17}$ \cite{GG}. STEP, using pairs of concentric free-failing 
proof-masses, will be able to test the WEP to a 
sensitivity at $1$ part in $10^{-18}$ \cite{Overduin}. 

On the other hand, 
notice that these mentioned experiments don't investigate the WEP 
in the case of charged particles. The reason is that electromagnetic fields 
influence gravitation experiments with charged particles and must be shielded 
carefully. The experiments for freely falling electrons carried out by 
Witteborn and Fairbank \cite{Witte}, with an accuracy of $10^{-1}$, is the 
only one cited in literature. Nowadays, Dittus and L\"ammerzahl \cite{Dila} 
showed that an experiment in space with the Witteborn-Fairbank set-up may 
be well suited to test the WEP and to improve the results for free fall 
test with charged particles by orders of magnitude.

Some important comments should be made on these two formulations of the
Equivalence Principle (EP). The SEP is valid only in static and
homogeneous gravitational field, but it is always possible
to choose a sufficiently small space-time region where
the gravitational field can be locally approximated by a
static homogeneous field, so that the SEP is valid locally.
For a scalar particle, the Pauli formulation of EP proposes
that a homogeneous gravitational field can always be transformed
away globally so that in a suitable reference frame there is only
Minkowski space - no gravitational field. On the other hand,
Audretsch \cite{Audretsch} observes that
if one takes a particle with spin, the equation of motion for
such a particle will inevitably involve the curvature tensor,
which can not be eliminated by any transformation of coordinates. 
Some authors ignore the influence of curvature
(second derivates) or tidal effects, but this means that they get rid of
gravitational field. Finally, some results has been obtained for an
infinite homogeneous gravitational
field (in the entire space) or for an uniformly accelerated boundless 
reference frame. These gravitational fields are not a true gravitational 
fields \cite{Rohrlich4}.

About the WEP, the equations of motion 
of a point mass in a curved background spacetime were investigated by 
Mino, Sasaki and Tanaka \cite{Mino}. 
The same equations of motion were later obtained by 
Quinn and Wald \cite{Quinn,Quinn2} from an axiomatic approach. Following 
Mino, Quinn and Wald, 
Haas and Poisson \cite{Poisson1} calculate the self-force acting on a point 
scalar charge in a wide class of cosmological spacetimes. The self-force 
produce two effects: a time-changing inertial mass and a deviation 
relative to geodesic motion. The work of Dewitt and
Brehme \cite{Dewitt}, corrected by Hobbs \cite{Hobbs}, showed that a
point charge in true gravitational field not follows a geodesic,
so that WEP is violated for a charged particle. Using the
techniques of finite-temperature field theory, Donoghue et al.
\cite{Donoghue} showed that the equality of inertial mass and
gravitational mass, for charged spin-$\frac{1}{2}$ or spin-zero
particles, is no valid in the context of quantum field theory at
finite temperature. Higuchi \cite {Higuchi} calculated the position shift 
of the final-state wave packet of the charged particle due the radiation 
and showed that it disagrees with the result obtained using the 
Lorentz-Abraham-Dirac equation for the radiation-reaction force. 
In an alternative approach, Spohn
\cite{Spohn} and other authors \cite{Blinder}-\cite{Rohrlich5} changed 
the Lorentz-Abraham-Dirac
equation for the force on an accelerating charge, which avoids
the pathologies of preacceleration and runaway solutions. 
Yaghjian \cite{Yaghjian} suggest that these problems will be absent 
once the finite-size effects are properly taken into account. Finally, 
the validity of the WEP is very well tested for macroscopic
bodies to a sensitivity of few parts in $10^{-13}$, 
but this does not necessarily imply that such principle
continues to hold at a microscopic scale and in the quantum
regime.

The goal of this
work is not to discuss about these papers, but, instead, to add the
possibility to analyze the problem of local motion of the classical 
charged point particles in a different perspective, with
emphasis in the EP, which validity is used as a good
starting point. As the subject of this work is about the conditions of
validity or not of the EP, it is
just to notice that results from General Relativity are not
used at any moment, in order to avoid any possibility to
fall in a vicious causal recurrence.

This text is a review of an old work of Goto \cite{Mariohep}.
What we have to do is to figure out the condition that
we have to provide such that a classical charged point particle can reach a 
locally stable hyperbolic motion. We show that is necessary to furnish a 
balance between an applied external force and the electromagnetic 
radiation reaction force to get an hyperbolic motion regime. An
important consequence is that, taking account the SEP, it
implies in a passive gravitational mass that is slight greater than the inertial mass.
From this result, one show that what seems to be
uncomfortable, as the presence of radiation for the charged particle performing
hyperbolic motion and its absence for one supported at rest in an uniform
gravitational field \cite{Fulton,Boulware}, both equivalent situations
as the SEP is valid, lead to a new
physical feature performed by charged particles. 
As consequence of the unexpected behavior of the passive gravitational and inertial 
mass relation, at a very strong gravitational field, we find the presence of 
a divergence that indicates a critical field value that signs the validity 
domain of the SEP. 

This paper is organized as follows.
In section II, we show that the locally external force necessary
to produce an hyperbolic motion in neutral particles
is smaller than the locally external force necessary to give the same
hyperbolic motion in classical charged particles.
In section III, we figure out that, to the SEP to be valid, the WEP is
violated for classical charged point particles in stable local hyperbolic 
motion regime. More, we show that there exists a critical, particle dependent, 
gravitational field value that signs the validity domain of the SEP.
The section IV is devoted to a final discussion and conclusions.

\section{Local hyperbolic motion of charged particles}

Hyperbolic motion is the natural generalization of the concept of the
Newtonian uniformly accelerated motion due to a constant force applied to a
particle, which might be due to an uniform gravitational field. At
relativistic level, as the velocity is upper limited by the light velocity,
constant force don't imply in constant acceleration; instead, it results in the
above mentioned hyperbolic motion, which denomination comes from the
hyperbola that it is drawn in the $zt$-plane by this kind of motion.

An one dimensional hyperbolic motion of a particle of mass $m$ occurs as 
a solution of the relativistic equation of motion \cite{Moller,Landau}

\begin{equation}
\label{eq_motion}
m\frac{d^{2}x^{\mu }}{d\tau ^{2}}    =f^{\mu }(\tau ) ,
\end{equation}
\vskip 0.2cm

\noindent
when external force ${F}$ is parallel to velocity ${v}$ and it is locally 
constant in the proper referential frame. In (\ref{eq_motion}) $f^{\mu }$
is the relativistic force defined as

\begin{equation}
\label{gama}
f^{\mu }=\gamma \left( \frac{{v\cdot F}}{c},{F}\right)
\hskip 0.5cm {\mbox{\rm with}} \hskip 0.5cm
\gamma =\frac{1}{\sqrt{1-\beta^{2}}}
\hskip 0.25cm {\mbox{\rm and}} \hskip 0.25cm
\beta=v/c \ ,
\end{equation}
\vskip 0.2cm

\noindent
where $c$ is the velocity of light.

Supposing the motion along the $z$-axes, the trajectory of hyperbolic motion
is given by

\begin{equation}
\label{orbit}
(z^{0},\rm{ }z)=\frac{c^{2}}{a}(\sinh \lambda \tau \rm{, }\cosh \lambda
\tau )\ ,
\end{equation}

\vskip 0.2cm

\noindent
where $a=F/m$ is a constant proper acceleration and $\lambda =a/c$. From
(\ref{orbit}) the velocity and acceleration are given by

\begin{equation}
(\stackrel{.}{z}^{0},\rm{ }\stackrel{.}{z})=c(\cosh \lambda \tau \rm{, }%
\sinh \lambda \tau )=\gamma c \, (1,\beta) \label{velocity}
\end{equation}
and
\begin{equation}
(\stackrel{..}{z}^{0},\rm{ }\stackrel{..}{z})=a(\sinh \lambda \tau \rm{,
}\cosh \lambda \tau )\ ,  \label{acceleration}
\end{equation}

\vskip 0.2cm

\noindent
respectively, so that the relativistic force responsible by the hyperbolic
motion is

\begin{equation}
f^{\mu }(\tau )=m(\stackrel{..}{z}^{0},\rm{ }\stackrel{..}{z})=ma(\sinh
\lambda \tau \rm{, }\cosh \lambda \tau )\ .  \label{force}
\end{equation}

\vskip 0.2cm

\noindent
The choice of the metric tensor $g^{\mu \nu }$ is such that $v^{\mu }v_{\mu
}=-c^{2}$ for four velocity $v^{\mu }=\stackrel{.}{x}^{\mu }$and, at non
relativistic limit, $a^{\mu }a_{\mu }=a^{2}$ for four acceleration $a^{\mu
}= $\/$\stackrel{..}{x}^{\mu }$.

The equation of motion of a classical charged point particle, including 
electromagnetic radiation reaction force, is given by the well known 
Lorentz-Abraham-Dirac equation 
\cite{Dirac}-\cite{Jackson}, \cite{Spohn}-\cite{Rohrlich5}, 

\begin{equation}
ma^{\mu }(\tau )=f_{ext}^{\mu }(\tau )+f_{rad}^{\mu }(\tau ) , \label{LD-eq}
\end{equation}

\vskip 0.2cm

\noindent
where $f_{ext}^{\mu }(\tau )$ is the external four-force and

\begin{equation}
f_{rad}^{\mu }(\tau )=m\tau _{0}\left( \stackrel{.}{a}^{\mu }-\frac{1}{c^{2}}%
a^{\nu }a_{\nu }v^{\mu }\right) ,  \label{LD-force}
\end{equation}

\vskip 0.2cm

\noindent
with

\begin{equation}
\tau _{0}=\frac{2}{3}\frac{e^{2}}{mc^{3}}\ ,  \label{tau-zero}
\end{equation}

\vskip 0.2cm

\noindent
is the Lorentz-Abraham-Dirac relativistic electromagnetic radiation reaction force.
The first term in
(\ref{LD-force}) is known as the Schott term \cite{Rohrlich} and it is
responsible by the well known non-physical runaway solutions. The second is
the Rohrlich term, related to the power radiated

\begin{equation}
{\cal R}=\frac{d\:{\rm{W_{rad}}}}{dt}=m\tau _{0}a^{\nu }a_{\nu }\ .
\label{rad-power}
\end{equation}
\vskip 0.2cm

A well known condition for hyperbolic motion, which satisfies
(\ref{orbit}-\ref{force}), is

\begin{equation}
\stackrel{.}{a}^{\mu }-\frac{1}{c^{2}}a^{\nu }a_{\nu }v^{\mu }=0\ ,
\label{hyperbole}
\end{equation}
\vskip 0.2cm

\noindent
which also implies in $f_{rad}^{\mu }(\tau )=0$, so it seems to be
easy to produce hyperbolic motion of a charged particle imposing a locally
constant external force, as in the uncharged particle case, but it
could induce to a misunderstanding. Hyperbolic motion is an ideal concept
that imply an eternal constant local acceleration, not existing in a real
world, and what happens immediately before reaching this regime will be
freezed in the final hyperbolic motion. The important result that we are going 
to show is that, while the final force that supports the hyperbolic motion for
uncharged and charged particles are equal, the composition of such forces is
different. For uncharged particles it is just the external force, but for
charged particles, it is composed by the sum of external force plus the
Rohrlich electromagnetic radiation reaction force. 
On other words, to have a stable 
hyperbolic motion for charged particles we have to get a very sensible 
balance between external and electromagnetic radiation reaction force, 
and before
it the condition (\ref{hyperbole}) is not true. It means that what happen 
before is
very important to get a stable hyperbolic motion regime and, although we
have the same Eq.(\ref{eq_motion}) after that, the force $f^{\mu
}(\tau )$ is not just $f_{ext}^{\mu }(\tau )$ anymore. To figure out why,
let us consider the Lorentz-Abraham-Dirac Eq.(\ref{LD-eq}) written as
\cite{Jackson}

\begin{equation}
m(1-\tau _{0}\frac{d}{d\tau })a^{\mu }=f_{ext}^{\mu }(\tau )-\frac{1}{c^{2}}%
{\cal R}v^{\mu }=K^{\mu }(\tau )\ .
\end{equation}

\vskip 0.2cm

\noindent
Formal expansion like

\begin{equation}
(1-\tau _{0}\frac{d}{d\tau })^{-1}=1+\tau _{0}\frac{d}{d\tau }+\tau _{0}^{2}%
\frac{d^{2}}{d\tau ^{2}}+\cdots  \label{formal}
\end{equation}

\vskip 0.2cm

\noindent
enables us to get a formal solution of Lorentz-Abraham-Dirac equation as

\begin{equation}
ma^{\mu }(\tau )=\sum_{n=0}^{\infty }\tau _{0}^{n}\frac{d^{n}}{d\tau ^{n}}%
K^{\mu }(\tau )\ .  \label{solution}
\end{equation}
\vskip 0.2cm

\noindent
We can insert the mathematical identity

\begin{equation}
\frac{1}{n!}\int_{0}^{\infty }s^{n}e^{-s}ds=1  \label{identity}
\end{equation}
\vskip 0.2cm

\noindent
to transform (\ref{solution}) in a second order integro-differential
equation

\begin{equation}
ma^{\mu }(\tau )=\int_{0}^{\infty }e^{-s}K^{\mu }(\tau +\tau _{0}s)ds\ ,
\label{ID-eq}
\end{equation}
\vskip 0.2cm

\noindent
which shows a possible non-causal behavior. In an explicit form, we have

\begin{equation}
ma^{\mu }(\tau )=\int_{0}^{\infty }\left. (f_{ext}^{\mu }-\frac{1}{c^{2}}%
{\cal R}v^{\mu })\right| _{\tau +\tau _{0}s}e^{-s}ds\ .  \label{non-causal2}
\end{equation}
\vskip 0.2cm

\noindent
The Eq.(\ref{LD-eq}), or the equivalent Eq.(\ref{non-causal2}),
are valid during the hyperbolic motion. On the other hand, 
while the motion is approaching the hyperbolic regime, as discussed above,  
we have the limiting process

\begin{equation}
\stackrel{.}{a}^{\mu }-\frac{1}{c^{2}}a^{\nu }a_{\nu }v^{\mu }\rightarrow
0\Rightarrow f_{rad}^{\mu }(\tau )\rightarrow 0\ ,  \label{limit}
\end{equation}
\vskip 0.2cm

\noindent
such that the total force behaves like

\begin{equation}
f_{ext}^{\mu }(\tau )+f_{rad}^{\mu }(\tau )\rightarrow f^{\mu }(\tau )\ ,
\label{limit2}
\end{equation}
\vskip 0.2cm

\noindent
or, $f_{ext}^{\mu}\rightarrow f^{\mu}$, so that the Eq.(\ref{non-causal2}),
in this regime, can be rewritten as

\begin{equation}
\int_{0}^{\infty }\left. (f^{\mu }-\frac{1}{c^{2}}{\cal R}v^{\mu })\right|
_{\tau +\tau _{0}s}e^{-s}ds=f^{\mu }(\tau )\ ,  \label{non-causal3}
\end{equation}
\vskip 0.2cm

\noindent
recovering the Eq.(\ref{eq_motion}), in accordance with Eq.(\ref{LD-eq}) 
and condition (\ref{hyperbole}).

From (\ref{non-causal3}), to have an hyperbolic motion, it is necessary that the applied external force goes to

\begin{equation}
f_{ext}^{\mu }(\tau )\rightarrow f_{ext}^{\mu }(\tau )=\int_{0}^{\infty
}f^{\mu }(\tau +\tau _{0}s)e^{-s}ds\ ,  \label{f-ext}
\end{equation}
\vskip 0.2cm

\noindent
as well as the electromagnetic radiation reaction force goes to

\begin{equation}
f_{rad}^{\mu }(\tau )\rightarrow f_{Roh}^{\mu }(\tau )=\left.
\int_{0}^{\infty }\frac{1}{c^{2}}{\cal R}v^{\mu }\right| _{\tau +\tau
_{0}s}e^{-s}ds\ ,  \label{f-Rohr}
\end{equation}
\vskip 0.2cm

\noindent
such that, from (\ref{limit2}-\ref{f-Rohr}),

\begin{equation}
f^{\mu }(\tau )=f_{ext}^{\mu }(\tau )+f_{Roh}^{\mu }(\tau )\ ,
\label{balance}
\end{equation}
\vskip 0.2cm

\noindent
where $f^{\mu }(\tau )$ is given by (\ref{force}).

Using (\ref{force}), the spatial component of Eq.(\ref{f-ext}) becomes

\begin{equation}
f_{ext}(\tau )=\frac{ma}{2}\left( \frac{e^{\lambda \tau }}{(1-\lambda \tau
_{0})}+\frac{e^{-\lambda \tau }}{(1+\lambda \tau _{0})}\right) \ .
\label{f-ext2}
\end{equation}
\vskip 0.2cm

\noindent
Analogously, from Eqs.(\ref{velocity}-\ref{force}) and (\ref{rad-power}),
the spatial component of Eq.(\ref{f-Rohr}) becomes

\begin{equation}
f_{Roh}(\tau )=-\frac{ma}{2}\lambda \tau _{0}\left( \frac{e^{\lambda \tau }}{%
(1-\lambda \tau _{0})}-\frac{e^{-\lambda \tau }}{(1+\lambda \tau _{0})}%
\right) \ .  \label{f-Rohr2}
\end{equation}
\vskip 0.2cm

\noindent
In the same way, time components become

\begin{equation}
f_{ext}^{0}(\tau )=\frac{ma}{2}\left( \frac{e^{\lambda \tau }}{(1-\lambda
\tau _{0})}-\frac{e^{-\lambda \tau }}{(1+\lambda \tau _{0})}\right)
\label{p-ext2}
\end{equation}

\noindent
and

\begin{equation}
f_{Roh}^{0}(\tau )=-\frac{ma}{2}\lambda \tau _{0}\left( \frac{e^{\lambda
\tau }}{(1-\lambda \tau _{0})}+\frac{e^{-\lambda \tau }}{(1+\lambda \tau
_{0})}\right) \ ,  \label{p-Rohr2}
\end{equation}
\vskip 0.2cm

\noindent
where $(1-\lambda \tau_{0})>0$ in (\ref{f-ext2}-\ref{p-Rohr2}). For
$(1-\lambda \tau_{0})<0$ the integrals (\ref{f-ext}-\ref{f-Rohr}) are
divergent. In section 3, this condition will be discussed.

From Eqs.(\ref{f-ext2}-\ref{p-Rohr2}), it is easy to see that the
total force (\ref{balance}) satisfies (\ref{force}), condition necessary
to have an hyperbolic motion. It shows that the external force necessary
to produce an hyperbolic motion in neutral particles,

\begin{equation}
f^{(neutral)}_{ext}(\tau )=ma\cosh \lambda \tau \ ,  \label{neutral}
\end{equation}
\vskip 0.2cm

\noindent
is smaller than the external force (\ref{f-ext2}) necessary to give the same
hyperbolic motion in charged particles. All external force applied to a neutral
particle is used to increase its kinetic energy,

\begin{equation}
\frac{d\:{\rm{W}}}{dt}=v\;F=m\;a\;v=m\;a\;c\;\sinh \lambda \tau =m\;c^{2}\;
\frac{d\:\gamma }{d\tau}\ ,  \label{pot}
\end{equation}
\vskip 0.2cm

\noindent
where $\gamma=\cosh \lambda \tau$ from (\ref{velocity}) and
$\lambda=a/c$. On the other hand, for charged particles, the
external force (\ref{f-ext2}), that can be written as

\begin{equation}
f^{(charged)}_{ext}(\tau )=f^{(neutral)}_{ext}(\tau )-f_{Roh}(\tau )\ ,
\end{equation}
\vskip 0.2cm

\noindent
provides the increase of kinetic energy in the same amount as given in (\ref
{pot}) and supplies, through $f_{Roh}(\tau )$, the energy lost carried by 
electromagnetic radiation.

\section{Classical charged particles in a local uniform 
gravitational field}

We saw in the previous section that a classical 
charged particle performing hyperbolic
motion have to be submitted to an external force given by

\begin{equation}
F_{ext}=\frac{ma}{2}\left( \frac{(1+\beta )}{(1-\lambda \tau _{0})}+\frac{%
(1-\beta )}{(1+\lambda \tau _{0})}\right) ,  \label{beta-force}
\end{equation}
\vskip 0.2cm

\noindent
where $F_{ext}$ is the spatial component of the measurable force related to
the relativistic force by $f_{ext}(\tau )=\gamma F_{ext}(\tau )$
[see (\ref{gama})]. To obtain (\ref{beta-force}) observe, from
(\ref{velocity}), that $\gamma(1+\beta)=e^{\lambda\tau}$ and
$\gamma(1-\beta)=e^{-\lambda\tau}$.

The SEP \cite{Rohrlich2,Rohrlich3} says
that a particle at rest in the laboratory frame $R_{lab}$ immersed in an
uniform gravitational field $g$ is seen by observer in a free falling
inertial frame $R_{in}$ as performing hyperbolic motion with local constant
acceleration $a=g$. The local constant force responsible by its hyperbolic
motion is the normal force $F_{n}=-F_{g}$ that supports the particle against
the gravitational force $F_{g}$, so that, in absolute value it is equal to $mg$
for a uncharged particle. But, for a charged particle, the normal force $F_{n}$
must be equal to the external force $F_{ext}$ of Eq.(\ref{beta-force})
for $\beta =0$,

\begin{equation}
F_{ext}\rightarrow F_{n}=\frac{mg}{(1-\lambda ^{2}\tau _{0}^{2})}\ .
\label{f-normal}
\end{equation}

\vskip 0.2cm

This result suggests that the observer in the laboratory frame $R_{lab}$
measures the gravitational force acting on a charged particle as

\begin{equation}
F_{g}=-\frac{mg}{(1-\lambda ^{2}\tau _{0}^{2})}\   \label{f-grav}
\end{equation}

\vskip 0.2cm

\noindent
such that

\begin{equation}
m^{\ast }=\frac{m}{(1-\lambda ^{2}\tau _{0}^{2})}\   \label{grav-mass}
\end{equation}

\vskip 0.2cm

\noindent
should define the passive gravitational mass $m^{\ast }$ of a charged particle 
with inertial mass $m$. The relation (\ref{grav-mass}) shows that, to the SEP 
to be valid, the WEP is violated for classical charged particles in stable 
hyperbolic motion regime.

On other words, the external force applied on uncharged
particles must be slightly lower than for charged particles. The first one
disposes all the external force to increase its kinetic energy, while the
charged one needs an external force to supply the same kinetic energy plus
the energy lost by electromagnetic radiation. As the strong version of the
EP is invoked, the equivalence between the proper
uniformly accelerated hyperbolic motion referential and the referential
supported in the presence of an uniform gravitational field implies that the
charged particle gravitational force, as well as the gravitational potential
energy, must be slightly greater than for the uncharged one. This difference
should be interpreted as due to their different gravitational masses and the
same inertial masses. It implies that the gravitational and inertial masses
of charged particles are different.

For typical charged particles, as electron or proton,
$\tau _{0}\simeq 6.3\times 10^{-24}s$ and $\tau _{0}\simeq 3.4\times
10^{-27}s$, respectively. In a field magnitude typical for a terrestrial
gravitational field $g\simeq 10 \; m/s^2$, we have 
$\lambda =g/c\simeq 3.3\times 10^{-8}s^{-1}$. So, we can see that 
$\lambda ^{2}\tau _{0}^{2}\simeq 4.\,3\times 10^{-62}$ and
$\lambda ^{2}\tau _{0}^{2}\simeq 1.3\times 10^{-68}$,
respectively for electron and proton, very small numbers, such that $%
1-\lambda ^{2}\tau _{0}^{2}\cong 1$. As a consequence, the passive gravitational
mass is just slight greater than the inertial mass, $m^{\ast }\gtrsim m$,
and the gravitational and inertial mass relation defined by Eq.(\ref
{grav-mass}) is much as close to unit, $r=m^{\ast }/m\cong 1$. It means
that there is no consequence, for practical purpose, due to this slight up
deviation of passive gravitational mass in relation to the inertial mass, 
at least 
in a region with gravitational field of magnitude as considered above. In
fact, it is true for a very long field interval, starting with $g=0$ and
going until to reach a very strong gravitational field of the order $%
g\thicksim 10^{30}m/s^{2}$.

In a such strong gravitational field region, the field dependence of 
$r=m^{\ast }/m\:$,
see Eq.(\ref{grav-mass}), starts to manifest, as we can see in the
figure 1. It increases very slowly and remains very close to unit, starting
with $g=0$ until to reach the very strong field magnitude of the order $%
g\thicksim 10^{30}m/s^{2}$, approaching the divergence point given by the
condition $1-\lambda ^{2}\tau _{0}^{2}=0$. Then, $r$ increases fast to
the infinite as the gravitational field goes to its critical value
$g_{c}=c/\tau _{0}$. This critical field value is mass dependent, with
$g_{c}\simeq 4.8\times 10^{31}m/s^{2}$ and $g_{c}\simeq 8.8\times
10^{34}m/s^{2}$ for electron and proton, respectively. 

The divergence of the gravitational and inertial mass relation
at the critical field value signs that the SEP
should not be valid in this situation. Consistently, the condition
$(1-\lambda \tau_{0})>0$ for which the integrals
(\ref{f-ext}-\ref{f-Rohr}) are finites, implies that
$(1-\lambda \tau_{0})(1+\lambda \tau_{0})=(1-\lambda ^{2}\tau _{0}^{2})>0$,
so that the investigation of the values above the critical
point $g_{c}$, as the mass relation $r$ turns to be negative, is nonsense.

Where is possible to find such strong gravitational field? 
Astrophysical compact objects are the natural place to search for more
intense gravitational fields. For example, white dwarf like Sirius B, 
with mass close to the solar mass, $1.0\times M_{\odot }$, 
and radius near $5.5\times 10^{6}m\simeq0.008R_{\odot }$, is a very compact 
object with a surface gravitational acceleration $g\simeq 4.6\times
10^{6}m/s^{2}$ {\cite{Barstow}}. (The solar mass and radius are $M_{\odot }\simeq
1.98844\times 10^{30}kg$ and $R_{\odot }\simeq 6.961\times 10^{8}m$,
respectively). 

Neutron stars are more compact than white dwarfs. For mass of order  
$1.4\times M_{\odot }$ corresponds a radius of order 
$R\simeq 4.4\times 10^{3}m$ with
a surface gravitational acceleration $g\simeq 2\times 10^{12}m/s^{2}$ 
{\cite{Carrol}}.

Extreme compact objects are of course the black holes. Caution is
need to deal with such objects, because there is no back information access
beyond their events horizon, defined by the spherical surface at the
Schwarzschild radius given by {\cite{Carrol}}

\[
R_{S}=\frac{2MG}{c^{2}}\simeq 2.95\times \frac{M}{M_{\odot }}km.
\]

\vskip 0.2cm

\noindent
The gravitational acceleration at the Schwarzschild surface is {\cite{Carrol}}

\[
g=\frac{MG}{R_{S}^{2}}=\frac{c^{4}}{4MG}\simeq 1.5\times 10^{13}
\frac{M_{\odot }}{M}m/s^{2}.
\]

\vskip 0.2cm

\noindent
For a typical black hole with about ten times solar mass, $M\simeq 10\times
M_{\odot }$, $R_{S}\simeq 30km$ and the gravitational acceleration at its
Schwarzschild surface is $g\simeq 1.5\times 10^{12}m/s^{2}$. It is belief
that galaxies central region shelter very massive black holes, with masses
of order $10^{5}\times M_{\odot }$ to $10^{9}\times M_{\odot }$. Such black
hole, with mass $M\simeq 10^{5}\times M_{\odot }$ has $R_{S}\simeq
2.95\times 10^{5}km$ and gravitational acceleration at its Schwarzschild
surface $g\simeq 1.5\times 10^{8}m/s^{2}$. If the mass is $M\simeq
10^{9}\times M_{\odot }$, the Schwarzschild radius is $R_{S}\simeq
2.95\times 10^{9}km$ and $g\simeq 1.5\times 10^{4}m/s^{2}.$ 
Increasing the black hole mass doesn't imply increasing the gravitational
acceleration, because the Schwarzschild radius increase together.  It shows that
astrophysical environment hardly get us gravitational acceleration stronger
than about $g\sim 10^{13}m/s^{2}$, which implies $\lambda ^{2}\tau
_{0}^{2}\sim 10^{-36}$, nothing to worry about. Only gravitational
acceleration strong as $g\sim 10^{30}m/s^{2}$ are going to be sensible in
equation (\ref{grav-mass}), close to an infinite singularity. Such too
strong gravitational field is very unlikely. 

On the other hand, collision of very energetic particles could take place with 
instantaneous acceleration that might be of such order, simulating a local and 
instantly gravitation, as that needed to test the equivalence principle. 
Nevertheless, it seems to be more
comfortable to abdicate of the Principle of Equivalence in such extreme
regime, where quantum effects certainly are dominant, and the gravitation, as it
is a belief, is going to be unified with the other interactions. 

\section{Conclusion}

Although the final equation of motion of classical neutral and charged 
particles in hyperbolic motion regime seems to be identical, we realize 
that there is a fundamental difference between them. For classical 
charged particles, actually, the total locally constant force is the sum 
of an applied and the electromagnetic radiation
reaction forces in a combination in such a way that results the same
as for neutral particles. An interesting implication, as the SEP
is taken account,
is that the passive gravitational mass of the charged particle must be greater
than its inertial mass in a very small amount. It is small enough to not be
detected by any experimental or practical devices, 
but it helps us to figure
out the condition necessary to get an hyperbolic motion regime and to
understand the meaning of its equivalence with a charged particle supported
at rest in an uniform gravitational field.

Until now the equality between gravitational and inertial mass was
understood as the essence of the EP, condition that we
realized not to be true for charged particles, since, due to the presence of
electromagnetic radiation reaction, a slight deviation of gravitational mass 
compared to inertial one is necessary to hold the SEP. However, a new
feature comes from the gravitational and inertial mass relation behavior for
a very strong gravitational field. There exists a critical,
particle dependent, gravitational field value that signs the validity domain 
of the SEP. For electron and proton these critical field 
values are $g_{c}\simeq 4.8\times 10^{31}m/s^{2}$ and $g_{c}\simeq 8.8\times
10^{34}m/s^{2}$, respectively. They possibly coincide where the quantum 
effects turns to be relevant.

\newpage
\centerline{\Large \bf{Figure Caption}}
\vskip 0.7cm

Figure 1 - Electron gravitational and inertial mass relation, 
$r=m^{\ast }/m$, as function 
of the gravitational field $g$. There is a critical point defined by
$g_{c}=c/\tau _{0}$, which is particle dependent, and has the value
$g_{c}\simeq 4.8\times 10^{31}m/s^{2}$ for electron.

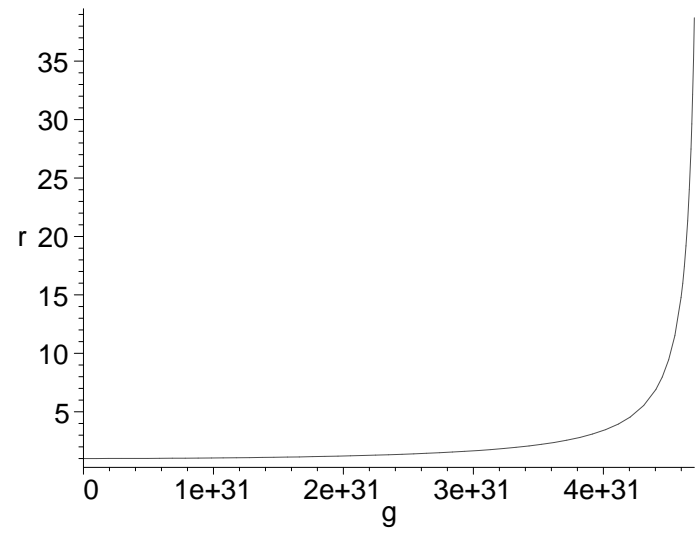
\end{document}